\documentclass[a4paper,11pt]{article}
\pdfoutput=1 

\usepackage{jinstpub} 

\usepackage{hyperref}
\hypersetup{colorlinks=true,
urlcolor=blue,
anchorcolor=blue,
citecolor=blue,
filecolor=blue,
linkcolor=blue,
menucolor=blue,
linktocpage=true,
pdfa=true
}

\usepackage[binary-units=true]{siunitx}
\DeclareSIUnit\torr{torr}
\graphicspath{{./figures/}}

\title{\boldmath Measurements of charging-up processes in THGEM-based particle detectors}

\author[a,1]{M. Pitt,\note{Corresponding author.}}
\author[b]{P. M. M. Correia,}
\author[a]{S. Bressler,}
\author[a]{A. E. C. Coimbra,}
\author[a]{D. Shaked Renous,}
\author[b]{C. D. R. Azevedo,}
\author[b]{J. F. C. A. Veloso}
\author[a]{and A. Breskin}

\affiliation[a]{Dept. of Astrophysics and Particle Physics, Weizmann Institute of Science,\\
 P.O. Box 26, Rehovot 76100, Israel\\}
\affiliation[b]{I3N -  Physics Department, University of Aveiro\\
Campus Universit\' ario de Santiago 3810-193, Aveiro, Portugal}

\emailAdd{michael.pitt@weizmann.ac.il}

\abstract{The time-dependent gain variation of detectors incorporating Thick Gas Electron Multipliers (THGEM) electrodes was studied in the context of charging-up processes of the electrode's insulating surfaces. An experimental study was performed to examine model-simulation results of the aforementioned phenomena, under various experimental conditions. The results indicate that in a stable detector's environment, the gain stabilization process is mainly affected by the charging-up of the detector's insulating surfaces caused by the avalanche charges. The charging-up is a transient effect, occurring during the detector's initial operation period; it does not affect its long-term operation. The experimental results are consistent with the outcome of model-simulations.}


\begin{document}
\maketitle
\flushbottom

\section{Introduction}\label{sec:intro}

The stability of the detector's gain over time is vital for reaching a stable performance of gas-avalanche detectors, manifested in detector efficiency and energy resolution. Changes in the detector gain over the first hours of operation have been observed in gaseous detectors incorporating insulating electrode substrates; examples are in Gas Electron Multipliers (GEM)~\cite{SAULI20162,4179872,4395404}, Large Electron Multiplier (LEM)~\cite{1748-0221-9-03-P03017,1748-0221-10-03-P03017}  and Thick Gaseous Electron Multiplier (THGEM) - based detectors~\cite{Chechik:2006jm,1748-0221-4-08-P08001,1748-0221-10-02-P02012,1748-0221-10-09-P09020,1748-0221-5-03-P03009,1748-0221-10-03-P03026}. The gain variation has been commonly attributed to charging-up of the detector's insulating surfaces that modifies the electric field in the charge-multiplication region. A phenomenological model was proposed for simulating these gain variations, in GEM~\cite{1748-0221-9-07-P07025} and recently in THGEM~\cite{Correia:2017qdx} detectors. The model of the charging-up process in THGEM-like detectors addresses several features related to the gain stability; its dependence on time, on electrode and detector geometry, and on the applied voltages.
In this work, an experimental study has been carried out in different THGEM-electrode configurations (e.g. electrode holes with and without etched rims), different irradiation conditions and different gases - to validate the model-simulation results~\cite{Correia:2017qdx}. The experimental setup and methodologies are detailed in section~\ref{sec:setup}; they are followed, by an extensive comparison of the measured and simulated gain-stabilization results, under different experimental conditions (section~\ref{sec:shortterm}) and a detailed study of the hole-rim effects (section~\ref{sec:longterm}). The results are concluded in section~\ref{sec:conclusions}.

\section{Experimental setup and methodology}\label{sec:setup}

\subsection{The detector setup}\label{subsec:detector}
Small (\SI{30x30}{\milli\metre}) \SI{0.4}{\milli\meter} thick FR4 THGEM electrodes, copper clad on both sides, were used in this work. The electrode holes of \SI{0.5}{\milli\meter} diameter, were arranged in an hexagonal lattice with a pitch of \SI{1}{\milli\meter}. The operation with dual-face THGEM electrodes with holes having \SI{0.1}{\milli\meter} wide etched rims was compared with that of holes without rim. 
A drift gap of \SI{5}{\milli\meter} between the cathode and the top electrode and an induction gap of \SI{2}{\milli\meter} between the bottom THGEM electrode and the readout anode were maintained. A drift field of \SI{0.2}{\kilo\volt\per\centi\meter} and an induction field of \SI{0.5}{\kilo\volt\per\centi\meter} were used, unless otherwise stated. 

\begin{figure}[htbp]
\centering 
\includegraphics[width=0.8\textwidth,origin=c,angle=0]{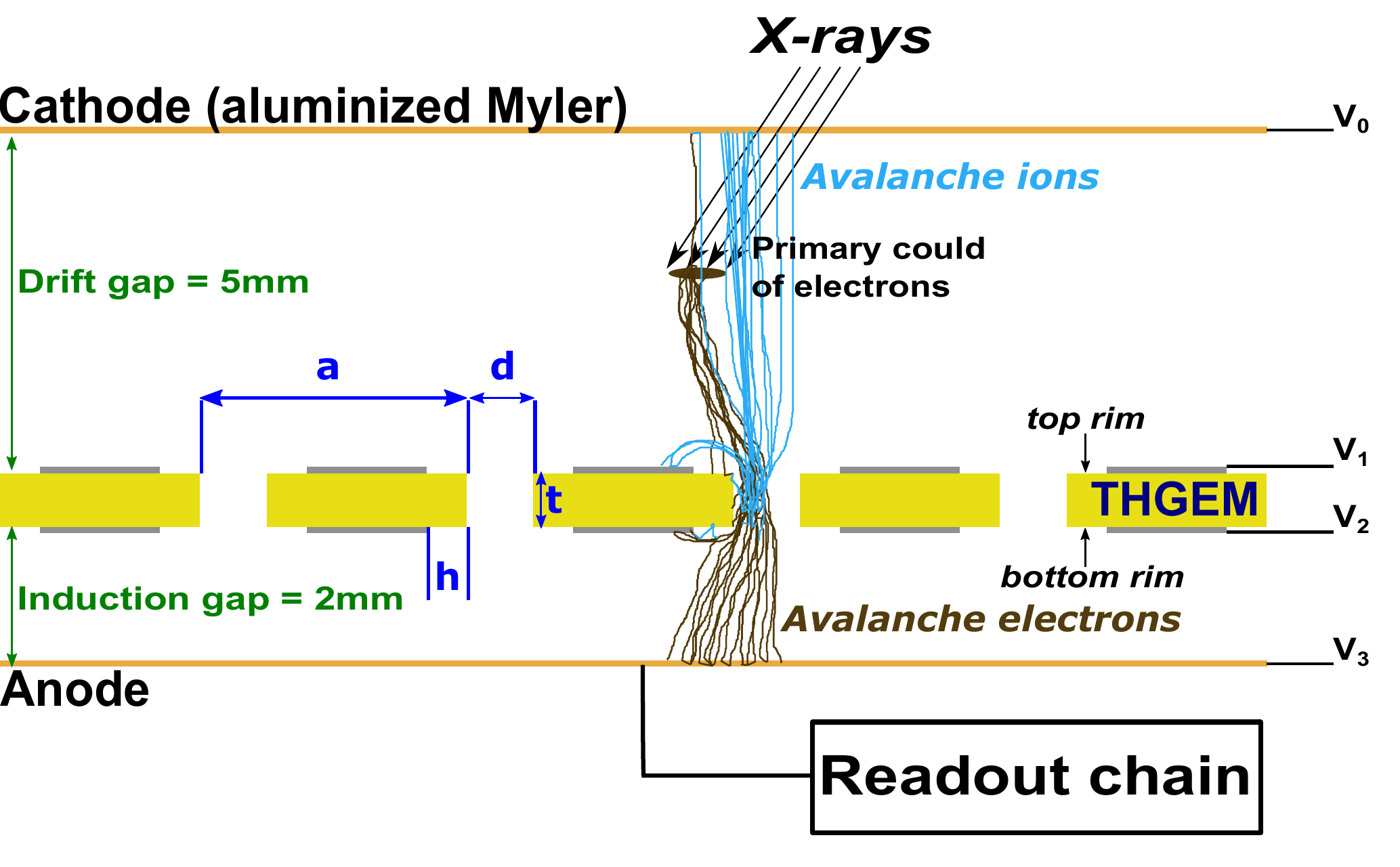}
\caption{\label{fig:scheme1} Scheme of the THGEM detector irradiated by an external $^{55}$Fe x-ray source; the aluminized-Mylar window, serving as a drift electrode (cathode); multiplied charges are collected through the induction gap between the THGEM bottom-electrode and the readout anode. The detector parameters are: t, thickness; d, hole diameter; a, hole pitch; h, rim size at the hole circumference.}
\end{figure}

The investigated detectors were placed in a small aluminum vessel. They were flushed with gas at a constant flow of \SI{20}{sccm}. Noble gases (Ar, Ne) were used in most of the measurements since they are not electro-negative; electro-negative gas mixtures can partially attach charges accumulating on the THGEM insulating surfaces, introducing additional dependency of the gain-stabilization process. The detectors were irradiated through a \SI{20}{\micro\meter} thick aluminized Mylar window, with \SI{5.9}{\kilo\electronvolt} photons emitted from a $^{55}$Fe source. The scheme of the detector layout is shown in figure~\ref{fig:scheme1}.
The detector electrodes (except the grounded drift cathode) were positively biased through low-pass filters, using CAEN N471A power supply units. The signals were recorded from the anode using an amplification chain comprising a Canberra2006 charge-sensitive preamplifier, Ortec572A linear amplifier and an Ampek 8000a Multi-Channel Analyzer (MCA). The amplification chain was calibrated prior to each measurement, using pulse-generator pulses through the preamplifier's test input.

\subsection{Detector initialization}\label{subsec:initialization}

As described in~\cite{1748-0221-12-09-P09036}, measurements of charging up processes are strongly affected by the charge history of the electrode. Thus, the electrodes must undergo an initialization procedure prior to each measurement. To clean up the electrode's history (i.e. to remove all previously-accumulated charges from the insulator), two approaches have been tested: rinsing with solvents, or flushing for $\sim$\SI{1}{\minute} with a jet of ionized nitrogen. Solvent-based cleaning was done by immersing the electrode in the solvent for $\sim$\SI{1}{\minute} and drying it at \SI{100}{\celsius} for $\sim$\SI{30}{\minute}. Ionized-nitrogen jet cleaning was performed with an anti-static gun (Electrostatics Inc. model 190M). 
\begin{figure}[htbp]
\centering
\includegraphics[width=1\textwidth,origin=c,angle=0]{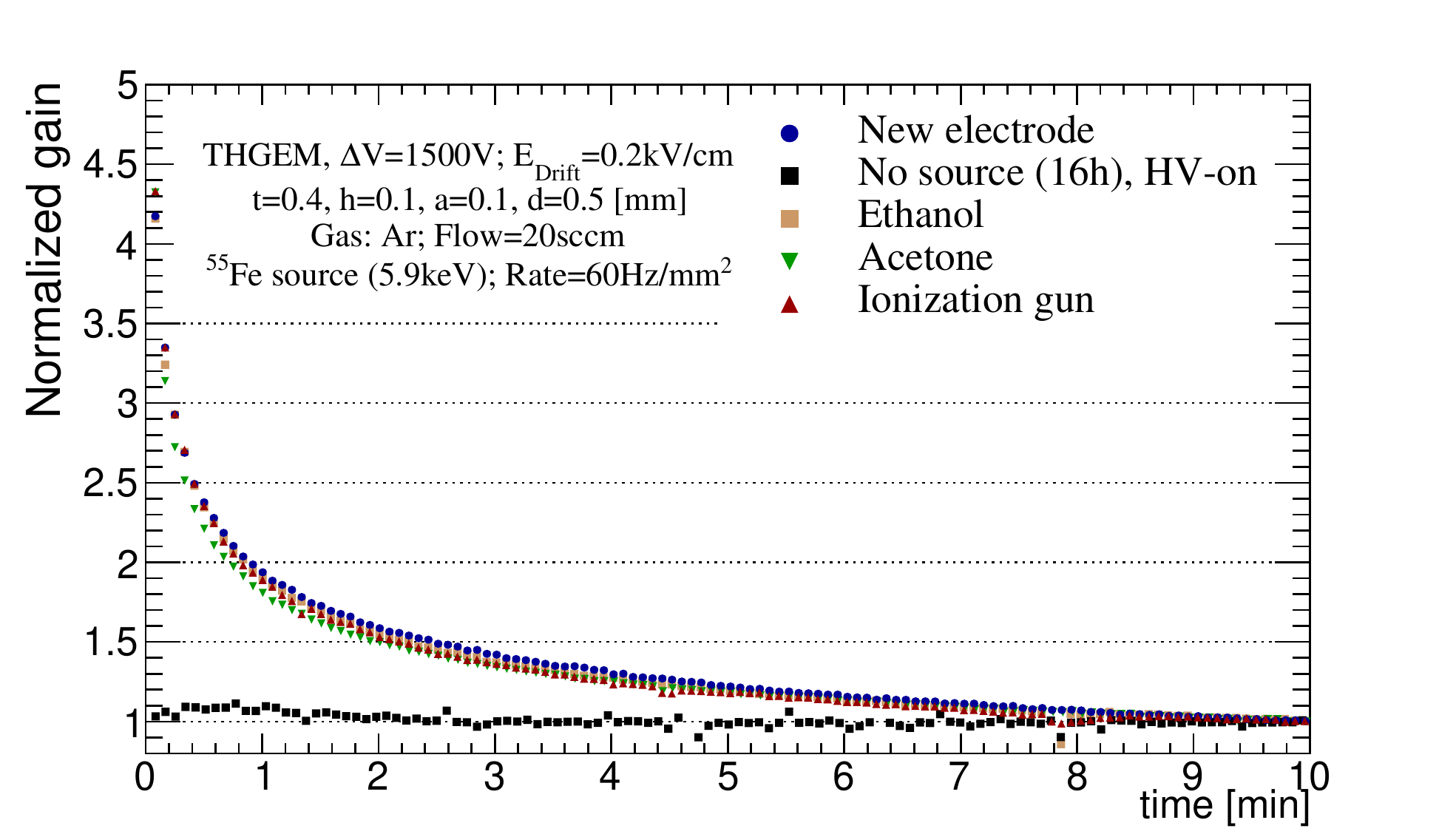}
\caption{\label{fig:initialization} Measured relative gain variation in Ar due to charging up of a new THGEM electrode, electrodes cleaned by different solvents or by ionized nitrogen; data are compared to an electrode having previous history, \SI{16}{\hour} after gain stabilization, with no irradiation but under applied voltage. The gain in each experiment was normalized to its value after 10 minutes of irradiation.}
\end{figure}
Results using these approaches are shown in figure~\ref{fig:initialization}, compared with that of a new electrode with no previous history. In addition, the gain stabilization was measured with an electrode that was irradiated until the gain was stabilized, followed by switching off the source for 16 hours while keeping the High-Voltage on; the gain was observed to remain constant, indicating that the accumulated charge remained on the insulator surfaces even under applied  voltage. The measurements were performed in Ar, to ensure no charge losses from detector's insulator by electro-negative gas molecules, and to avoid possible neutralization of the charged electrodes by the gas molecules.

\subsection{Determination of the detector working points}\label{subsec:workingpoints}
\begin{figure}[htbp]
\centering
\includegraphics[width=0.8\textwidth,origin=c,angle=0]{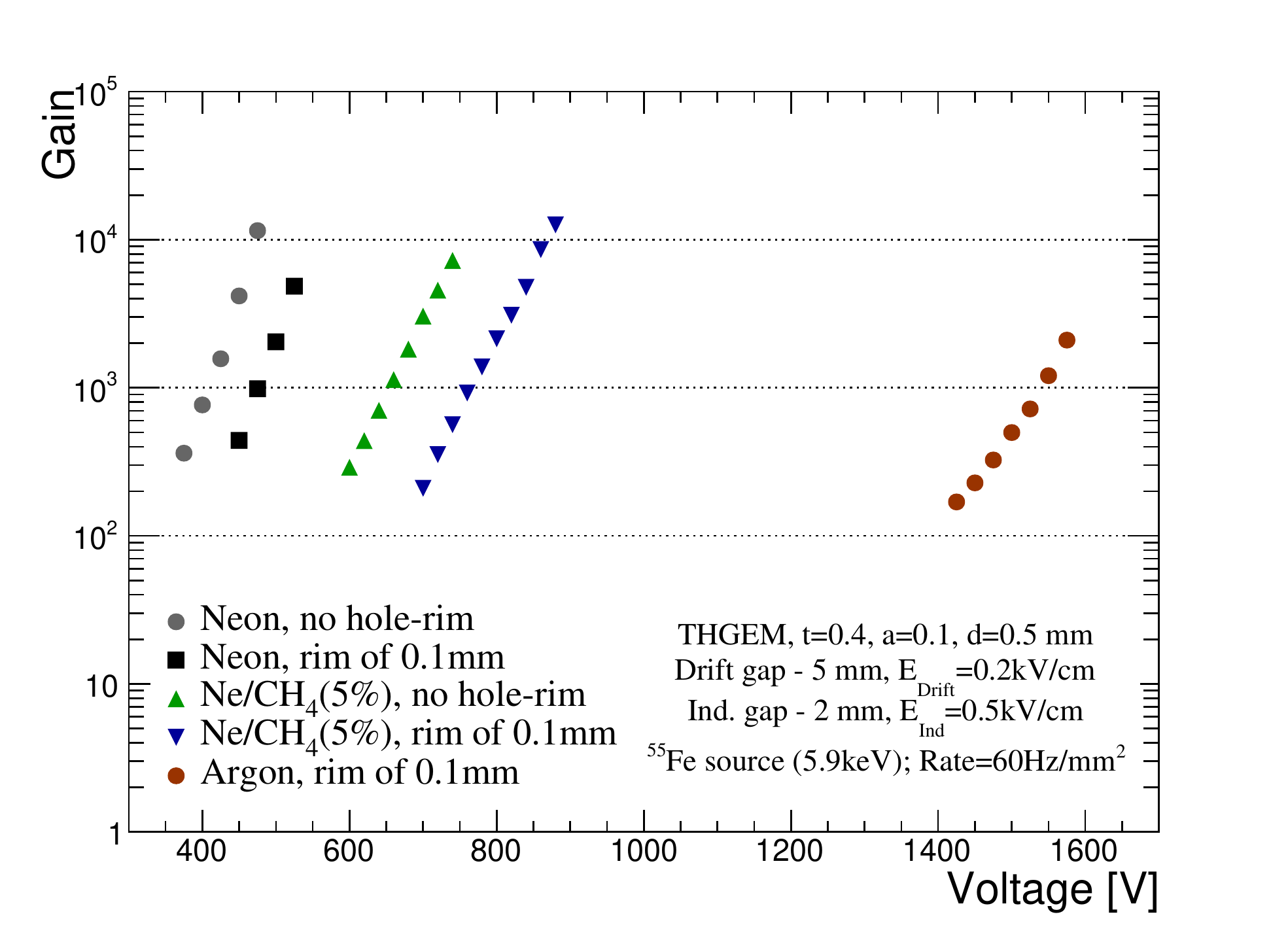}
\caption{\label{fig:workingpoints} Measured gain curves in different gases, in the THGEM detector, with \SI{5.9}{\kilo\electronvolt} X-rays, at a rate of \SI{60}{\hertz\per\milli\metre\squared} over a \SI{5}{\milli\meter} in diameter spot size.}
\end{figure}

Gain stabilization measurements were performed in different gases, this allows investigating detector operation over a broad range of voltages. It was observed, that above the highest achievable gain, an occasional discharge would cause the gain to increase. This observation can be explained by neutralization of accumulated charges during the discharge development. The effect of the discharge on the measured gain was not studied here. Therefore, to avoid potential discharge-induced effects, all measured gain values were recorded below the discharge limit. Typical gain curves for different gas mixtures studied here are shown in figure~\ref{fig:workingpoints}.
Gain measurements were performed at \SI{1}{\bar}, at room temperature, with a nominal gas flow of \SI{20}{sccm}. They were irradiated with $^{55}$Fe \SI{5.9}{\kilo\electronvolt} X-rays at a rate of \SI{60}{\hertz\per\milli\metre\squared}. The gain curves ended at voltage values where the first discharge occurred within the first five minutes of measurement. The detector was irradiated until reaching gain stabilization. Given these results, the operation voltages were fixed in a range between \SIrange{1450}{1550}{\volt} in Ar, \SI{740}{\volt} in Ne/CH4(5\%), \SIrange{475}{500}{\volt} in Ne for THGEM configuration with hole-rims, and \SI{425}{\volt} in Ne for a configuration without hole-rims.

\subsection{Simulation of charging-up processes}\label{subsec:simulation}
\begin{figure}[htbp]
\centering
\includegraphics[width=0.8\textwidth,origin=c,angle=0]{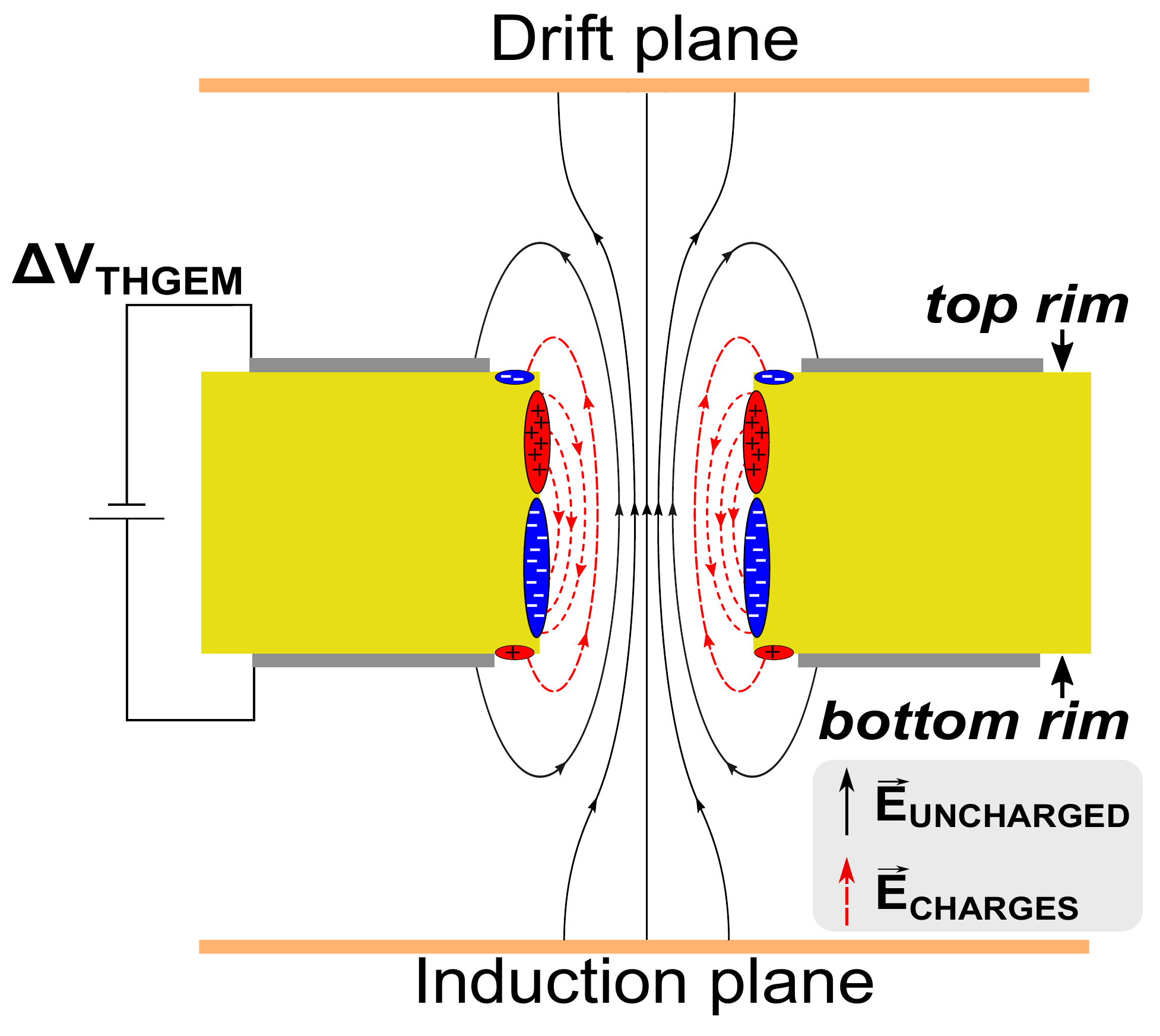}
\caption{\label{fig:superposition} Schematic drawing of charging-up of the THGEM electrode. The effective electric field map is a sum of the initial electric field, calculated with a potential $\Delta V_\text{THGEM}$ applied across the detector's electrode, superposed with the electric field induced by accumulated charges on the insulating surfaces of the detector.}
\end{figure}

The simulations of the charging-up of the detector's electrode, and the calculations of the avalanche-gain variations were performed using the recently developed simulation tool for THGEM-based detectors~\cite{Correia:2017qdx}. It consists of an iterative algorithm and is based on the superposition principle ~\cite{1748-0221-9-07-P07025}. The simulation toolkit evaluates electric-field variation due to accumulated charges on the insulating surface of the electrode. In each iteration of the algorithm, the total amount of the charge attached to the electrode hole's walls (or rim surface, as discussed below) is calculated, and the resulting electric field induced by the accumulated charge is superposed with the initial electric-field due to the potential applied across the electrode. This iterative procedure is repeated until the simulated charges are no longer accumulated on the insulating surface, and the total electric-field remaining constant with increasing number of iterations. A schematic drawing of the superposition principle is shown in figure~\ref{fig:superposition}.
The number of simulated iterations  ($n_\text{iter}$) is related to the physical irradiation time by:
\begin{equation}
\label{eq:timeiter}
t\left[\text{sec}\right]=\frac{s}{n_\text{p}\times R\left[\text{Hz}\right]}\times n_\text{iter}
\end{equation}
Where $n_\text{p}$ is the number of primary charges induced in the gas by the ionizing event, $R$ is the irradiation rate and  $s$, the step-size, is a constant value that multiplies the total number of accumulated charges on the THGEM electrode's walls in a given iteration. It corresponds to the number of avalanches considered in each iteration before changing the field configuration. (see~\cite{Correia:2017qdx} for more details).

\begin{figure}[htbp]
\centering
\includegraphics[width=0.8\textwidth,origin=c,angle=0]{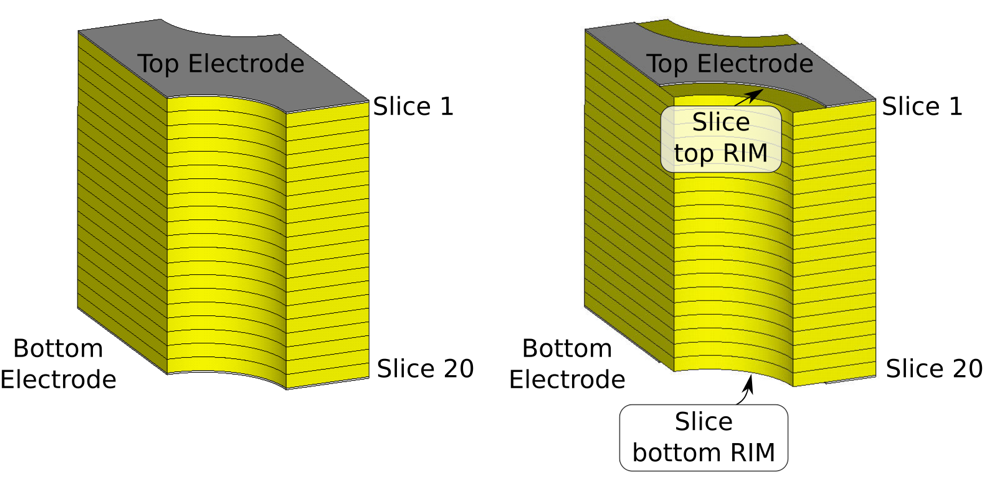}
\caption{\label{fig:cell} The THGEM cell employed in the simulations. Left: THGEM holes with no rim; the insulator surface is divided into 20 slices along the hole. Right: THGEM with hole rims; the insulator surface is divided into 20 slices + 2 surfaces.}
\end{figure}

In the simulation setup, THGEM electrode was divided into 20 slices, for a THGEM electrode with hole rims, the rims were included by introducing two additional slices - ``\textit{top rim}'' and ``\textit{bottom rim}'', as shown in figure~\ref{fig:cell}.

\subsection{Measurements of charging-up processes}\label{subsec:measurements}
In all the measurements, the detector electrode was initialized as described in section~\ref{subsec:initialization} using the anti-static gun. The chamber was flushed for at least one hour with a flow of \SI{20}{sccm}. The voltages were set one hour preceding irradiation, to ensure that the insulator is fully polarized. The gain variation over time was recorded with the MCA, in steps of 5 or \SI{30}{\second} depending on the irradiation rates. At each step, the measured pulse-height spectra were fitted to a Gaussian, the mean and the standard deviation values were recorded.

During the irradiation, the detector gain dropped due to accumulation of charges on the surface of the insulator walls. The characteristic stabilization time ($\tau$) is defined to be:
\begin{equation}
\tau=\tau_{0.5}/\log2
\end{equation}
where $\tau_{0.5}$ is the time when the gain decreased to half of the total gain variation. Such definition is made in order to relate the measured characteristic time to the one extracted from the simulation.
\subsubsection{Fit model}
The model of the gain stabilization in the simulations described by an exponential behavior as it was shown in~\cite{Correia:2017qdx}. The gain stabilization over time can be fitted to an exponential curve:
\begin{equation}
\label{eq:exp1fit}
G\left(t\right)=G_F+\Delta G \exp{\left(-t/\tau\right)}
\end{equation}
where $G_F$ is the initial gain, $\Delta G$ is the total gain variation and $\tau$ is the relaxation time constant (or ``\textit{characteristic time}''). Yet, the exponential curve is just a first approximation for the selected parameters in the simulation (drift and induction field). The actual gain stabilization curve is affected by the rate of the charge accumulation on the detector's walls, which might be different for different points along the hole's walls. As it will be shown later, drift and the induction fields affects detector's gain evolution. It was also observed that rates of accumulating charges on the surface of the hole's walls vary with the position of the charge accumulation (or slice in case of simulation), and the gain evolution in some cases might not match a single exponential function. It was observed, that in order to achieve a proper modeling of the short-term gain evolution the exponential curve must be extended. The short-term evolution of the measured gain over time was fitted to empirical fit function given by:
\begin{equation}
\label{eq:exp2fit}
G_\text{SHORT}\left(t\right)=G_F+\Delta G_1 \exp{\left(-t/\tau_1\right)}+\Delta G_2 \exp{\left(-t/\tau_2\right)}
\end{equation}
where $G_F$ is the final gain, $\Delta G_i$ is a gain drop value with corresponding ``\textit{characteristic time}'' $\tau_i$. This function was found to better model the gain evolution than single exponential curve, or other empirical functions provided elsewhere~\cite{1748-0221-9-03-P03017,1748-0221-10-03-P03017}.

The long-term gain evolution as it will be discussed later is a result of charge accumulation on a detector's ``\textit{top rim}'' (figure~\ref{fig:superposition}), such process will typically have a single characteristic time, and it is expected to be fitted to an exponential curve. The long-term evolution of the measured gain over time is fitted to an exponential function given by:
\begin{equation}
G_\text{LONG}\left(t\right)=\Delta G_\text{UP} \exp{\left(-t/\tau_\text{UP}\right)}
\end{equation}
Therefore, when the long-term component is present, the gain evolution curve is fitted to a linear combination of short- and long-term components:
\begin{equation}
\label{eq:exp3fit}
G(t)=G_\text{SHORT}+G_\text{LONG}
\end{equation}

In both simulation and measurements, gain stabilization is characterized in terms of the total amount of charge (in units of $q_e$) passing through the THGEM holes during relaxation period:
\begin{equation}
\label{eq:qtot}
Q_\text{tot}=q_e\times n_p\times R \int\limits_{0}^{\tau}G(t)dt
\end{equation}

\section{Evolution of the gain stabilization over short-term periods}\label{sec:shortterm}

\subsection{Comparison to simulation}\label{subsec:compare}
A comparison of simulated and measured gain stabilization was made for Ne/CH4(5\%). This mixture, of a known Penning transfer rate~\cite{1748-0221-11-08-P08018}, permits good comparison with the simulation results for a short-term gain stabilization time.
\begin{figure}[htbp]
\centering
\includegraphics[width=0.8\textwidth,origin=c,angle=0]{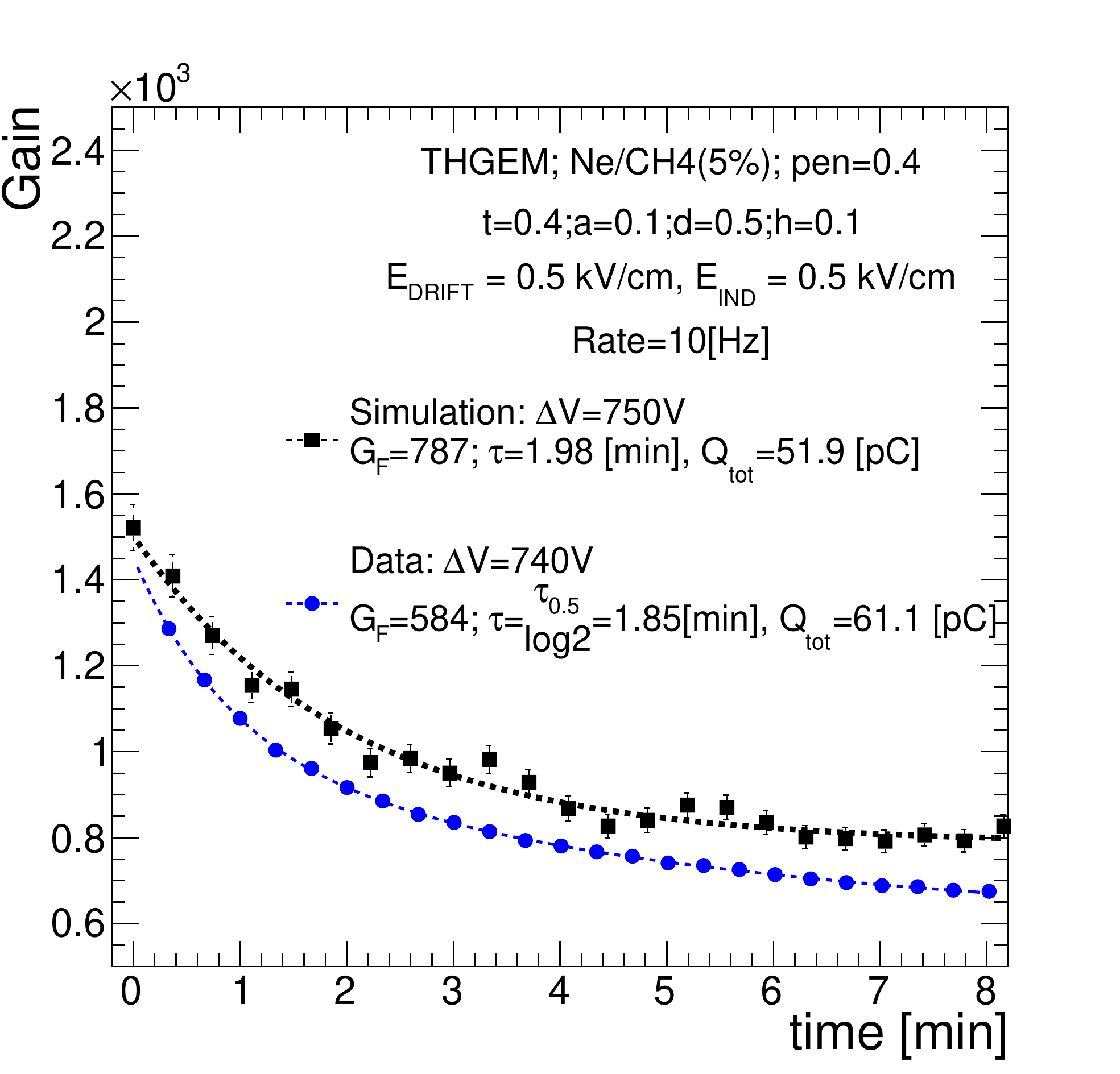}
\caption{\label{fig:simcompare} Gain evolution for an initial gain of $\sim$1500 vs. time in a THGEM operated in Ne/CH4(5\%). Simulation results (black) and measurements (blue). The measurements were performed for an electrode of clean history. The measured gain variation was fitted to an empirical function~\eqref{eq:exp2fit}, while the simulated one was fitted to a single-exponential function~\eqref{eq:exp1fit}. The parameters are shown in the figure.}
\end{figure}
Figure~\ref{fig:simcompare} shows a comparison between the simulated and measured gain evolution, over the first 8 minutes. In the measurements, the applied voltages were chosen such that the differences to an initial gain from the simulation are below 10\%. The stabilization time in both, simulation (black squares) and the measured data (blue circles) was found to be similar, approximately 2 minutes. The ``characteristic charge'' of the measured gain stabilization was calculated from the empirical fit and was found to be in a good agreement with the simulation result. To first approximation, the shape of the measured gain evolution with time was found to be consistent with simulations, and hence with the assumption that charging up of the THGEM hole dominates the gain-stabilization processes over this time. The difference in the shape between the data and the simulation results (different fit functions) could have several origins. In the measured data, some systematic variations in the experimental conditions may cause changes in the stabilization curve; different position of the x-ray spot over the area above the detector electrode, inhomogeneity in the electrode's geometry (variations in pitch, hole diameter, rim size, insulator thickness, etc.) or small amounts of gas impurities can also affect the gain stabilization. In the simulation, selected conditions as the number of slices or position of the initial charge deposition can also be a source of systematic variations of the results. In this work, we will examine qualitatively the charging up phenomena, pointing out at the major sources affecting the detector's stability. 
\subsection{Irradiation-rate effects}\label{subsec:irradiation}
To validate the argument that the source of gain variations is due to charging up of the insulator, measurements were carried out at different rates.
\begin{figure}[htbp]
\centering
\includegraphics[width=1.0\textwidth,origin=c,angle=0]{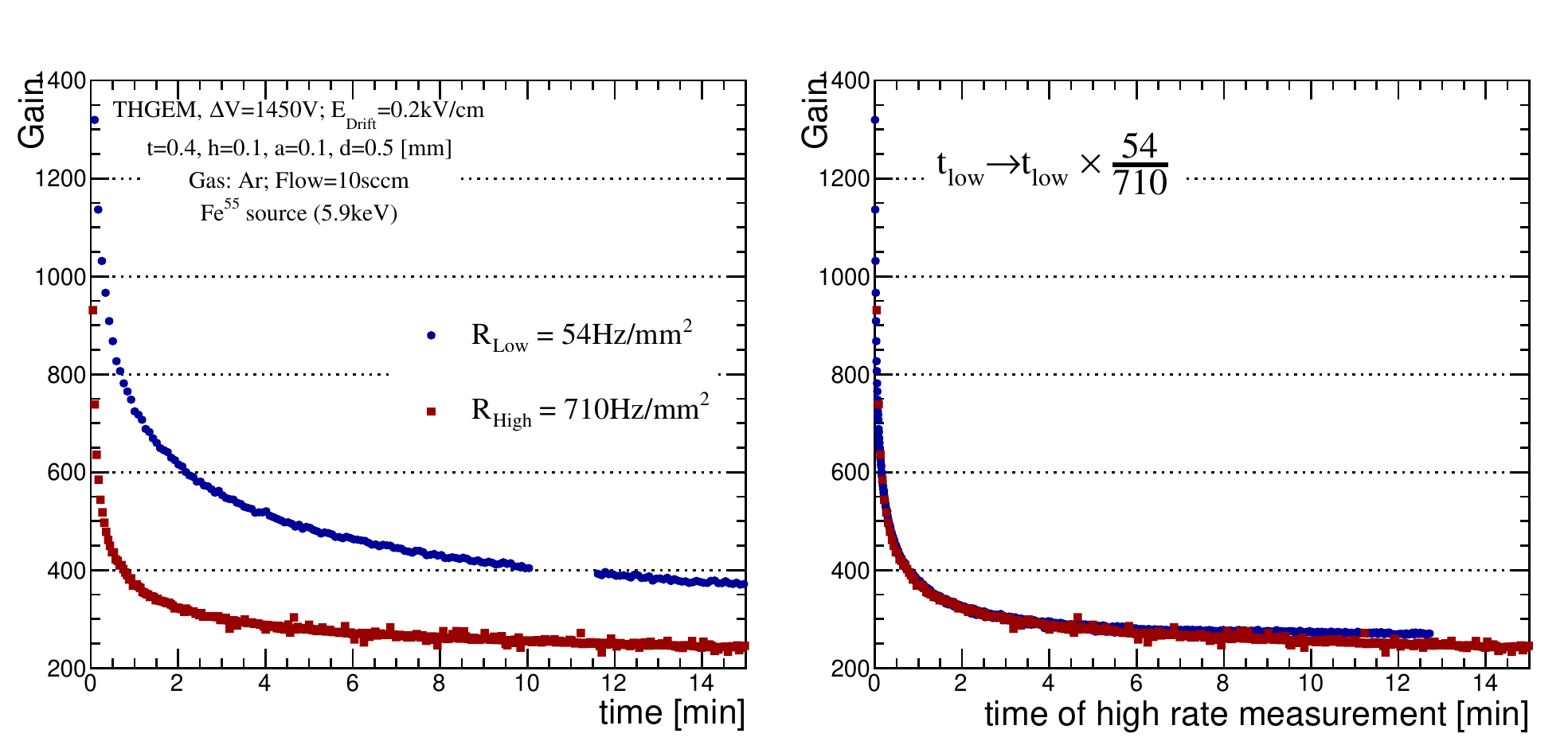}
\caption{\label{fig:rate} Measured gain stabilization (in Ar) for two different rates. In the left plot, gain stabilization is shown as a function of time, while in the right one, the time was scaled by the ratio of rates to compare the trends ($t\to t\times\frac{R_\text{low}}{R_\text{high}}$).}
\end{figure}
Figure~\ref{fig:rate} shows gain variation over time for different rates, measured in Ar. In the left plot of figure~\ref{fig:rate} the gain stabilization is shown as a function of time. The stabilization time is expected to be linear with the inverse of the irradiation rate~\eqref{eq:timeiter}. To show this effect for the two curves in the left plot of figure~\ref{fig:rate}, the x-axis (time) for the low-rate gain stabilization measurement was scaled by the ratio between the different rates, i.e. instead of $G_\text{low-rate}(t)$ we show $G_\text{low-rate}(t\times\frac{R_\text{low}}{R_\text{high}})$, and the two stabilization curves are compared on the right plot of figure~\ref{fig:rate}. 

\subsection{Charging up at different gains }\label{subsec:gains}
The short-term stabilization time at different gas gains was simulated and compared to measurements. The characteristic charge - the total charge that pass through the THGEM holes until the gain stabilizes - is proportional to the detector's gain, irradiation rate and the relaxation time~\eqref{eq:timeiter}. At higher detector gain, the charge that flows through the THGEM holes is higher; thus, the gain stabilization time is expected to occur faster. The electrode thickness in the simulations was chosen to be t = \SI{0.8}{\milli\meter} thick. The simulations were performed in Ne/CH$_4$(5\%) due to slightly faster execution time of an avalanche for this gas mixture.
\begin{figure}[htbp]
\centering
\includegraphics[width=1.0\textwidth,origin=c,angle=0]{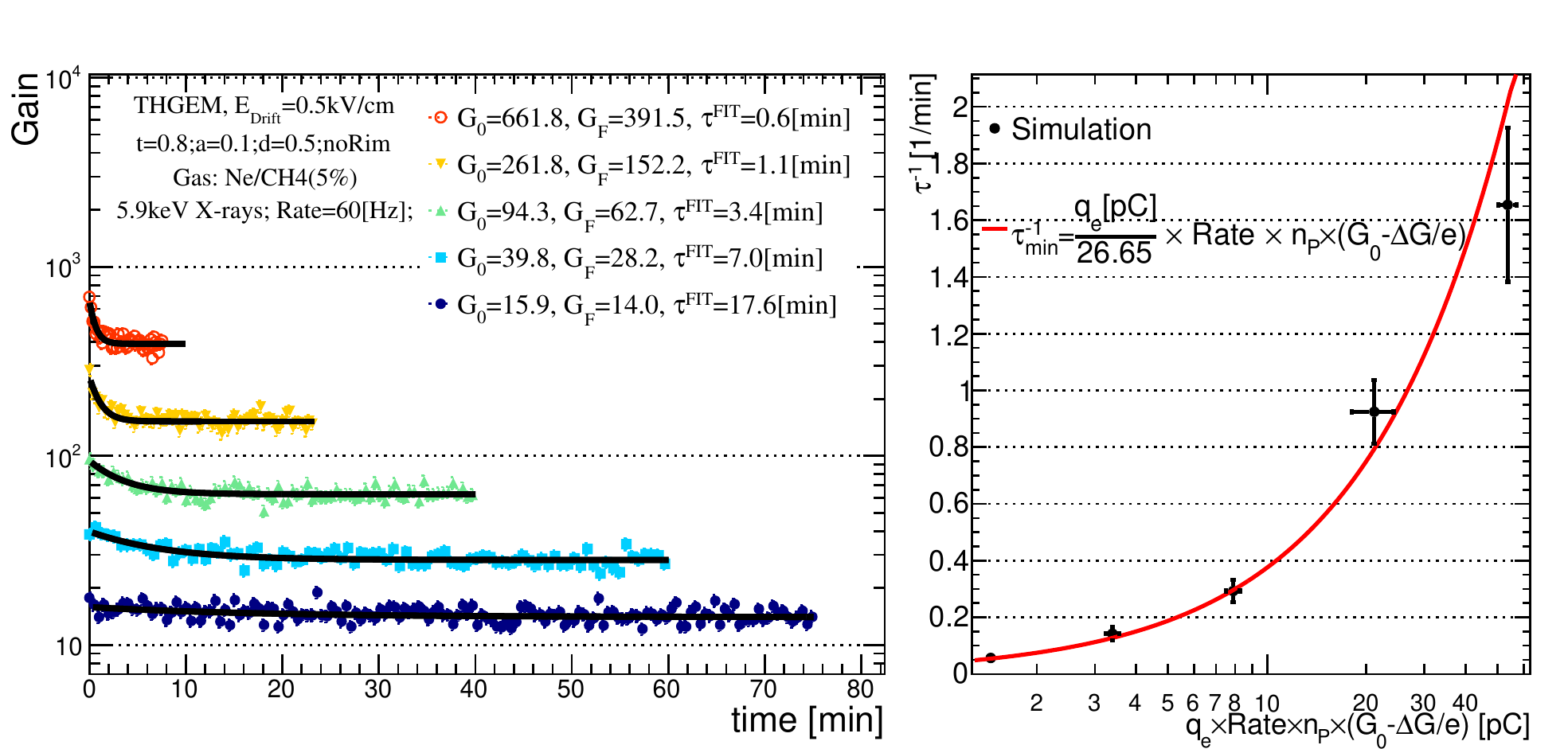}
\caption{\label{fig:gains_sim} Left: Simulated gain stabilization for different initial voltages (initial gains) as a function of time. The gain evolution is fitted by an exponential function; the fitted ``characteristic time'' values are shown in the legend. Right: Linear dependence (linear fit) of the inverse of the characteristic time as a function of the product of the gain and the irradiation rate.}
\end{figure}
Figure~\ref{fig:gains_sim}, shows the gain evolution for different applied voltages. From the fitted characteristic time and the initial gain, using equation~\eqref{eq:qtot}, one can obtain the $Q_\text{tot}$. The linear relation between the inverse of the characteristic time and the product of irradiation rate and the gain is also shown on figure~\ref{fig:gains_sim}. The difference in the applied voltages on the detector electrodes has negligible effect on the fitted $Q_\text{tot}$ value.

\begin{figure}[htbp]
\centering
\includegraphics[width=1.0\textwidth,origin=c,angle=0]{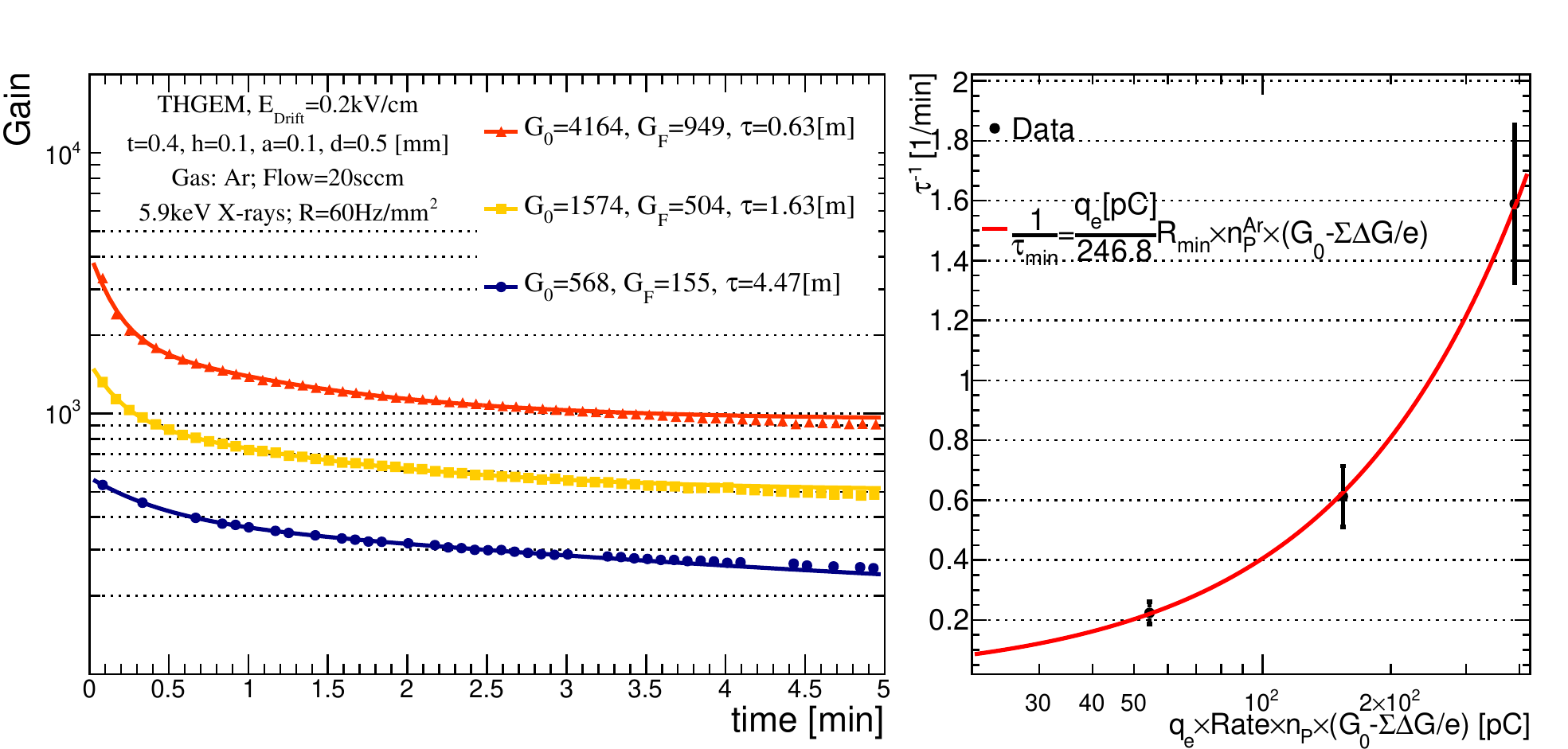}
\caption{\label{fig:gains_measure} Left: Measured gain stabilization (in Ar) over the first \SI{5}{\minute}, for different gain values. The characteristic time derived from the plots is defined to be the time where $G(\tau)=(G_0+G_F)/2$. Right: Linear fit of the inverse time with the product of the gain by the rate.}
\end{figure}

The stabilization time at different gains was measured in pure Ar (due to its larger ``characteristic time'' value than in Ne~\cite{Correia:2017qdx}). The electrode was irradiated with \SI{5.9}{\kilo\electronvolt} X-rays at a rate of \SI{60}{\hertz\per\milli\metre\squared}. Gain stabilization over time, for different gains, is shown in figure~\ref{fig:gains_measure}.
The characteristic time shows linear dependence with the initial gain - as suggested by the simulations. In both simulation and the measurement, the characteristic time is proportional to the inverse of the detector's gain. The ``characteristic charge'' estimated from the measured ``characteristic time'' vs the detector's gain, is also found consistent with that simulated for Ar-based mixtures (of order of hundreds of pC)~\cite{Correia:2017qdx}.

\section{Long-term time evolution of the gain - the effect of hole rims}\label{sec:longterm}
Previous studies showed that etching a small area around the holes (a rim; typically 0.05 -- 0.1\SI{}{\milli\metre}) reduces the discharge probability~\cite{1748-0221-4-08-P08001}. Although the useof rims has been shown to introduce gain variations along time~\cite{1748-0221-10-03-P03026}. As discussed in~\cite{Correia:2017qdx}, in THGEM detectors, charges originating from the avalanche that are attached to the hole's wall, reduce the electric field within the hole, resulting in a decrease of the gain over time. In the opposite, charges that are attached to the rim surface, are expect to increase the electric field in the detector holes since they have opposite charge to the charges attached to the hole's wall, resulting in an increase of the detector gain.

\subsection{Simulation of the rim effect on charging up }
\label{sec:simultion_slices}
To test the effect of the charging up as a position along the hole's insulator surface, only a single slice was allowed to charge up, while all others remained un-charged. After being fully charged (with no more electrons or ions attached to this slice), the final gain was estimated.
\begin{figure}[htbp]
\centering
\includegraphics[width=1\textwidth,origin=c,angle=0]{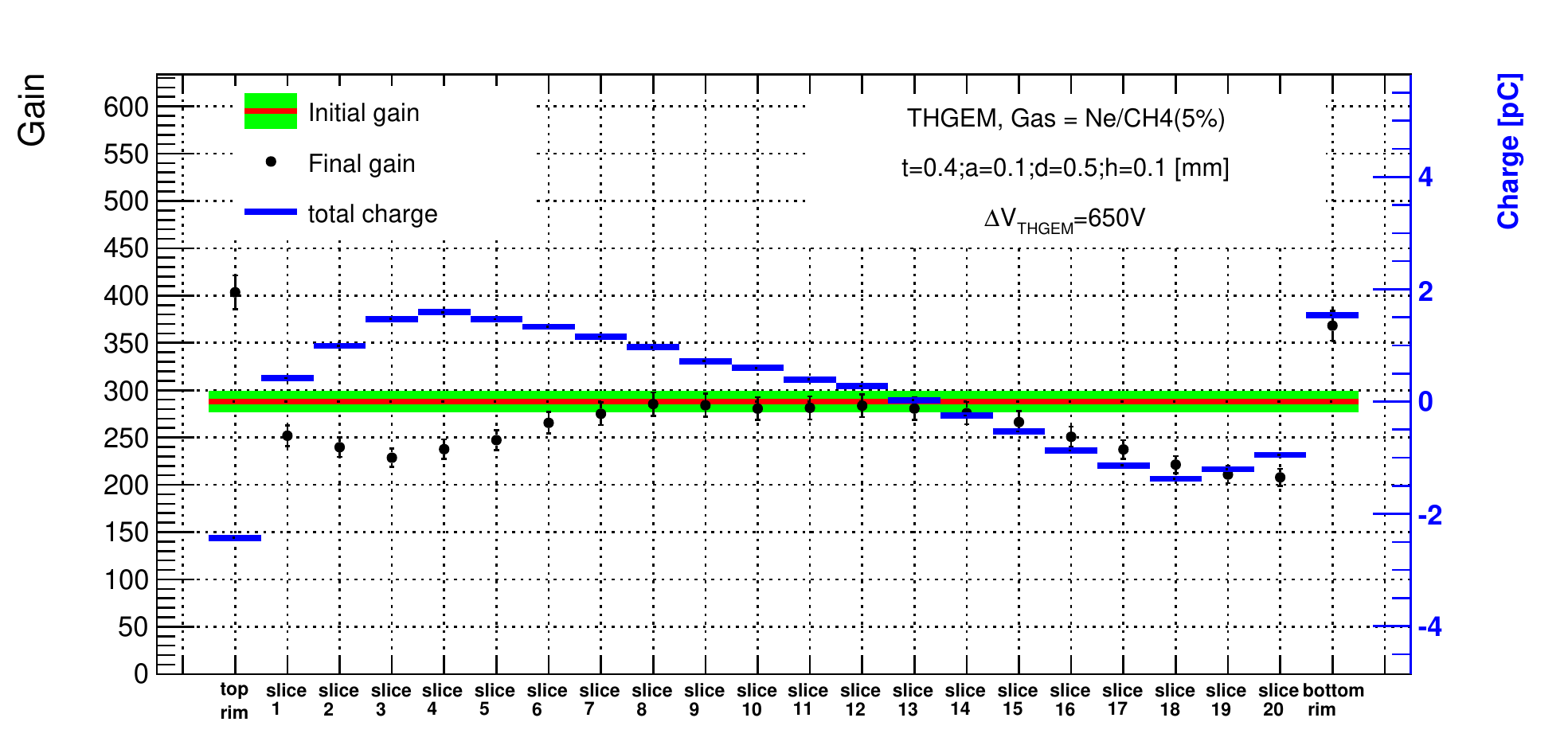}
\caption{\label{fig:slices} Simulated effect on the gain, resulting from each slice, and the distribution of accumulated charge along the detector hole. Shown are the initial detector gain (green band), the gain per slice when only one slice is charged-up (black) and the total charge accumulated on a single slice when detector gain reached stability (blue).}
\end{figure}
Figure~\ref{fig:slices} show the charge distribution on the insulating walls of a THGEM electrode with a \SI{0.1}{\milli\metre} rim, after gain stabilization. The black points show the simulated gain after stabilization, while the blue lines show the total charge accumulated on the insulator surface. As it is seen in the simulation result, the upper part of the hole's wall is positively charged (due to drifting ions from the avalanche process), and the lower half part of the hole's wall is negatively charged (avalanche electrons). These electrons and positive ions reduce the total gain, by creating an electric field in an opposite direction to that of the initial one.

The charges accumulate on the detector's hole rims in the following patterns: the ``\textit{bottom rim}'' is charged positively; drifting ions created in an avalanche close to the ``\textit{bottom rim}'' have a non-zero probability to deviate from the field lines that are focusing them into the holes, and positively charge the rim. The ``\textit{top rim}'' is charged negatively due to primary charges that were not focused to the detector holes. The characteristic charging-up time of the ``\textit{top rim}'' is therefore proportional to the initial number of charges (number of primary electrons induced by incoming ionizing event), while the characteristic charging-up time for the holes and the bottom rim is proportional to the final number of charges created during the multiplication process (initial number of charges multiplied by the detector gain).
\begin{figure}[htbp]
\centering
\includegraphics[width=0.3\textwidth,origin=c,angle=0]{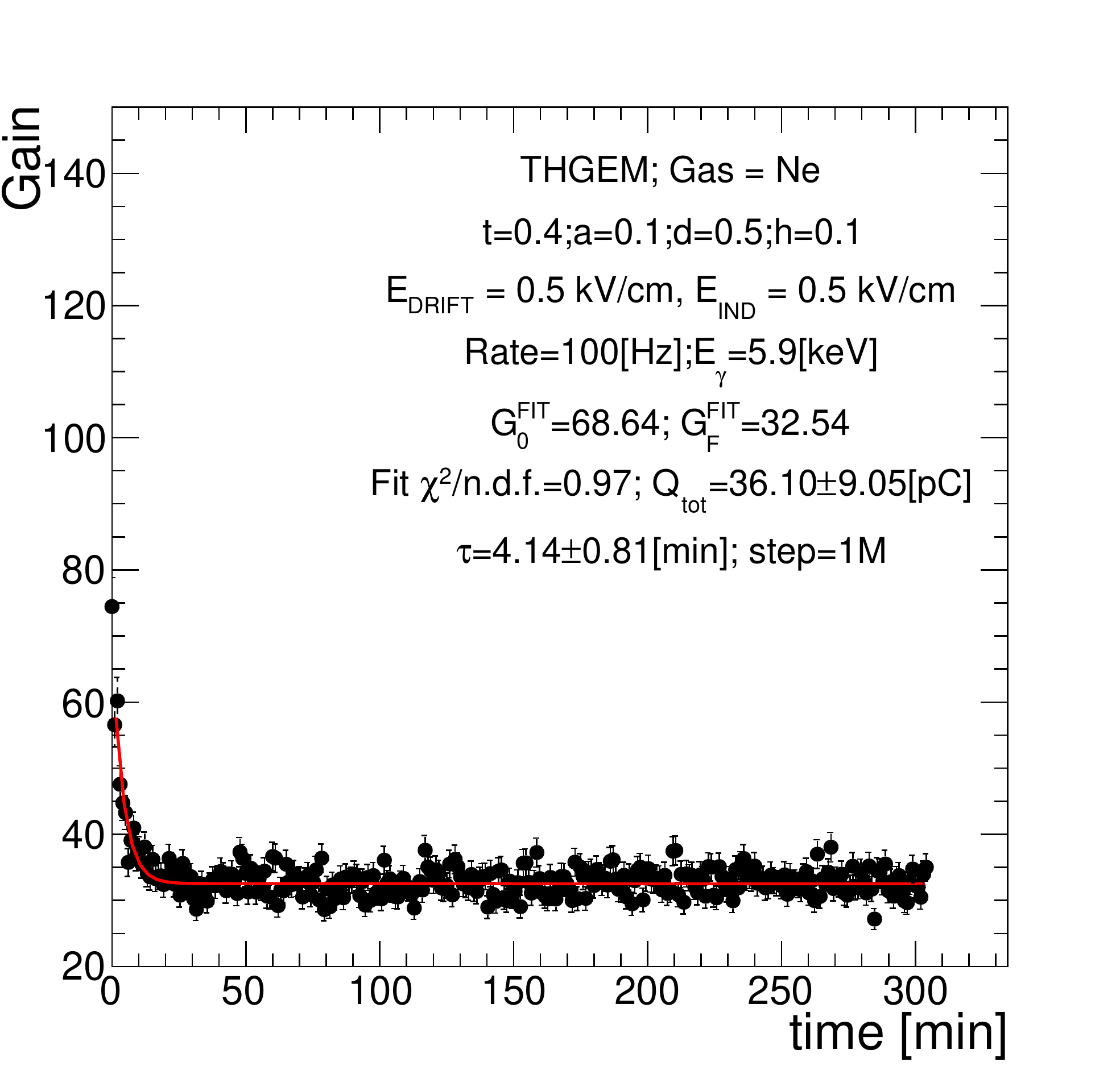}
\includegraphics[width=0.3\textwidth,origin=c,angle=0]{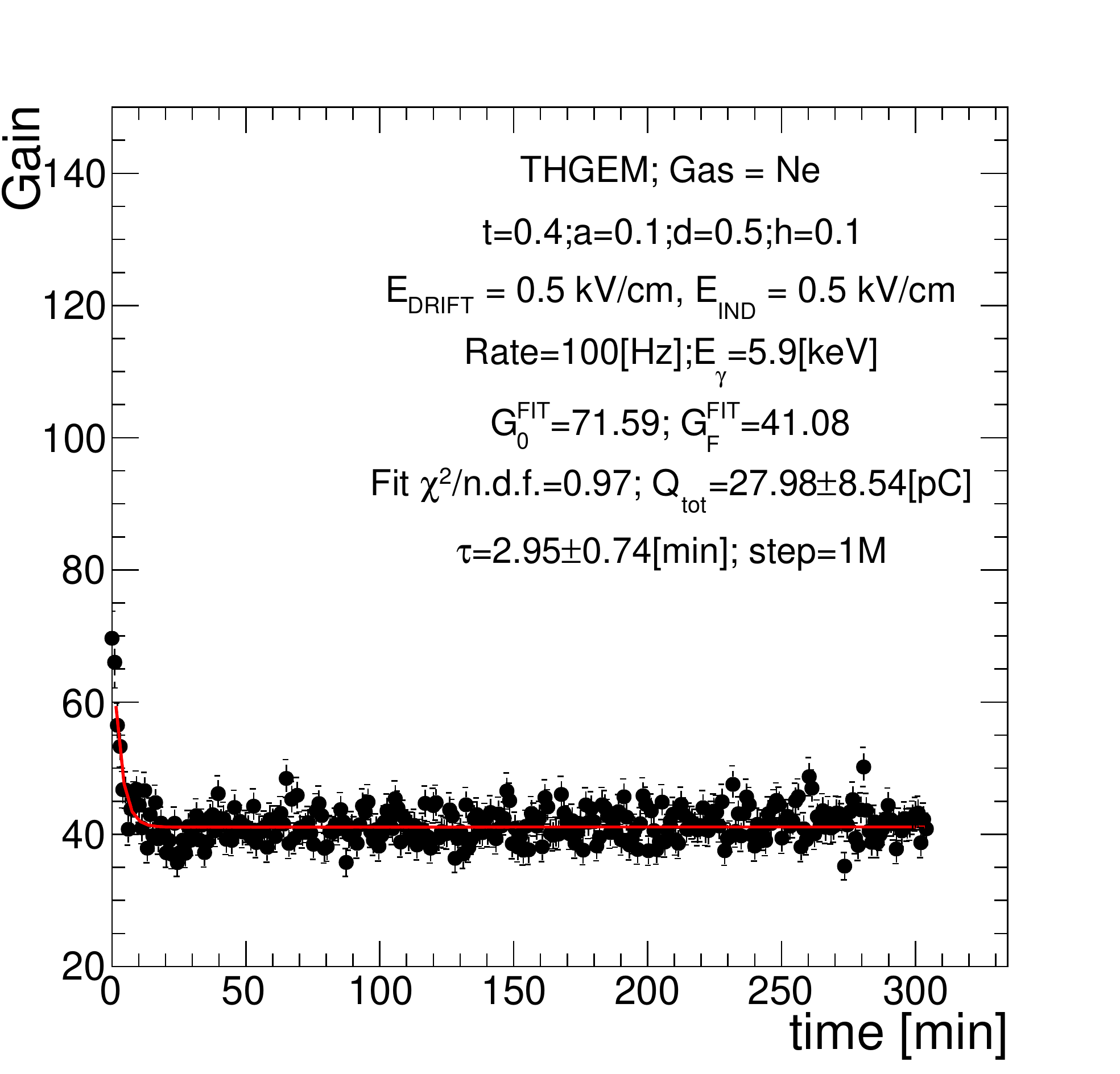}
\includegraphics[width=0.3\textwidth,origin=c,angle=0]{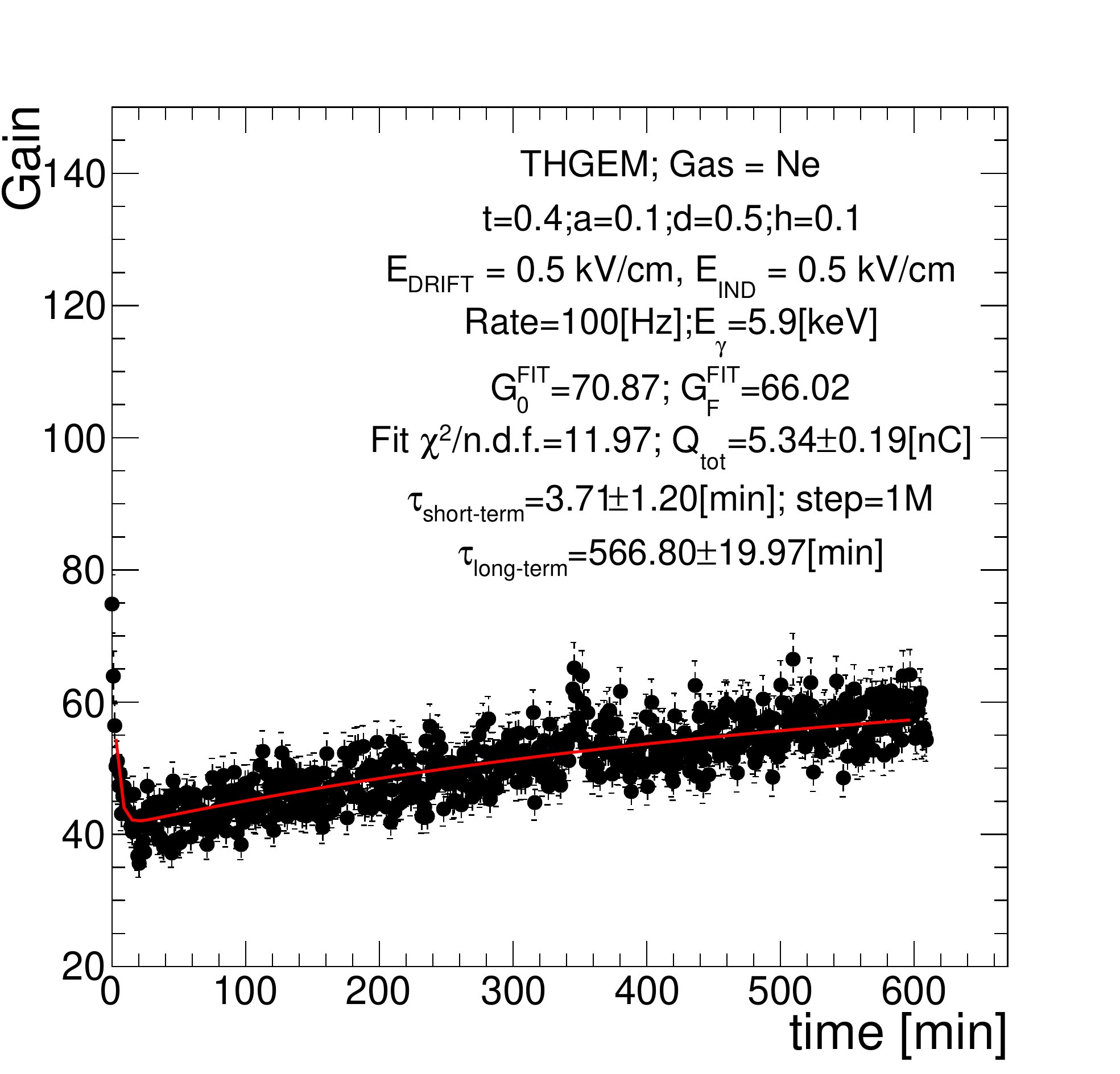}
\caption{\label{fig:rimapart} Simulated gain stability (in Ne) for an electrode with a rim of \SI{0.1}{\milli\metre}, for three different cases -- Left: none of the rims was charged up. Middle: top rim was not charged up. Right: all slices were charged up. The red lines are fit results of the simulation to an exponential function (for long-term component a double-exponential function is fitted).}
\end{figure}
Figure~\ref{fig:rimapart} shows the simulation results in pure Ne, with a rim of \SI{0.1}{\milli\metre} for three cases -- Left: charges are not allowed to accumulate on the top and bottom rims. Middle: charges are not allowed to accumulate on the top rim. Right: charges are accumulated on all slices. Since the effective time to charge up the bottom electrode, is similar to that for charging up the detector hole walls, the result of adding a bottom rim is the smaller gain drop. Charging up also the ``\textit{top rim}'', due to the accumulation of the charges from the primary electron cloud that are not focused to the THGEM holes, results in a long-term gain variation. Therefore, the difference between short-term and long-term variations is proportional to the total gain, and the electron transfer efficiency (ETE).

\subsection{Measurements of the rim effect on charging up}

\begin{figure}[htbp]
\centering
\includegraphics[width=0.8\textwidth,origin=c,angle=0]{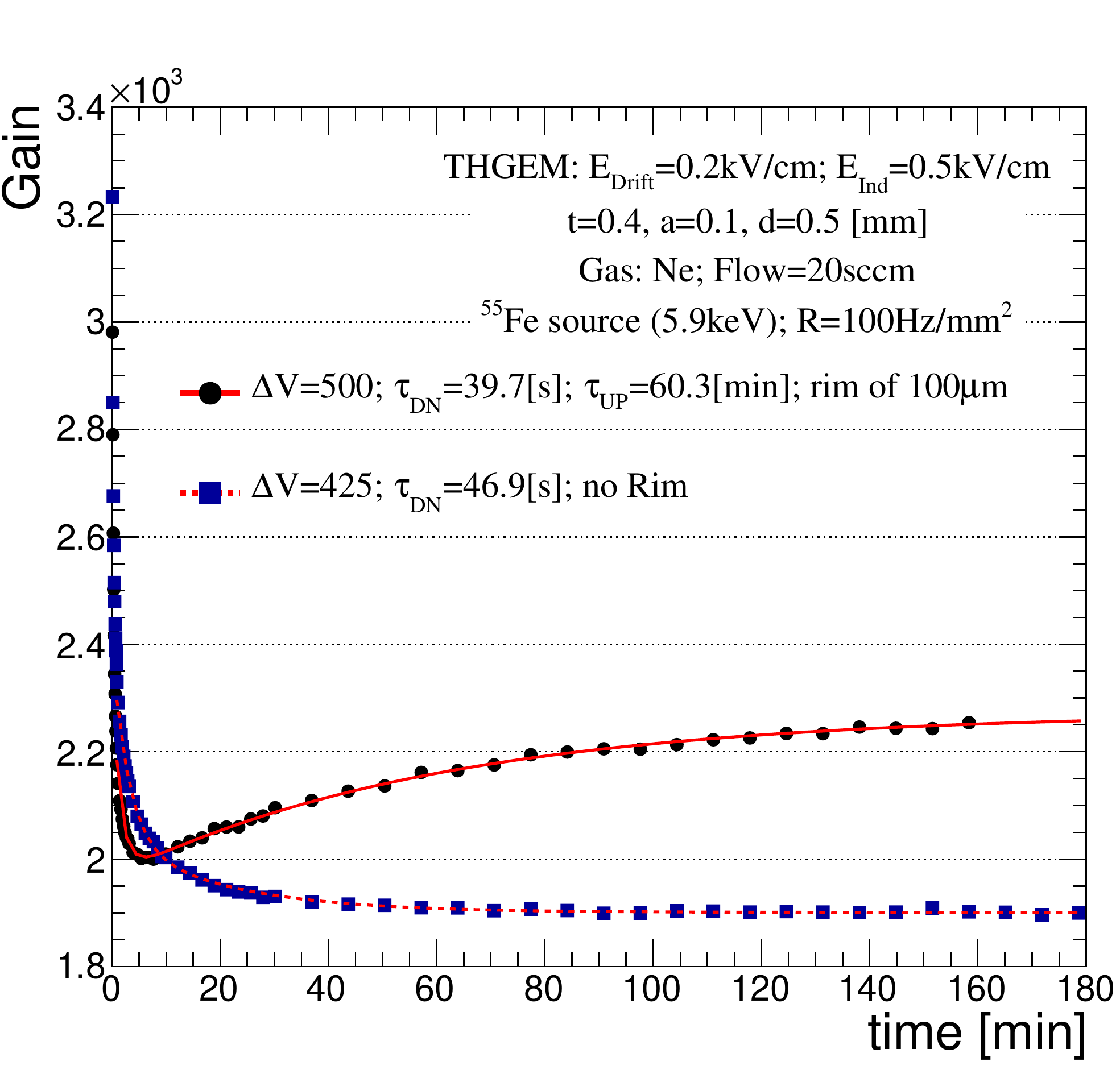}
\caption{\label{fig:rimeffect} Measured gain stabilization over time in Ne, at an X-ray rate of \SI{100}{\hertz\per\milli\metre\squared} for an electrode with hole-rim (black) and without rim (blue). The red lines are fit results of the measurement to a triple-exponential function~\eqref{eq:exp3fit} or double-exponential function~\eqref{eq:exp2fit} respectively.}
\end{figure}

The presence of hole-rims introduced long-term gain stabilization characteristic times, attributed to the long time scale of charging up the top rim with primary electrons. The effect of the rim was investigated with a THGEM electrode, in Ne (similarly to the simulation results in the previous section); the result of gain stabilization over time is shown in figure~\ref{fig:rimeffect}.
The gain variation over time, in case of an electrode with a ``\textit{top rim}'', was fitted to a function from equation~\eqref{eq:exp3fit}. In the THGEM configuration without a hole-rim, the gain is monotonically decreases with time, until it reach stable gain after about ten minutes. In the THGEM configuration with a hole-rim, we observe two characteristic timescales: a short one (minutes) and a long one (about few hours). The short timescale is related to a decrease of the detector's gain, the total gain drop measured for an electrode with a hole-rim is considerably smaller than for the configuration without a hole-rim, since an additional charging-up of the bottom rim increase of the gain. The rate of the charging-up of the ``\textit{bottom rim}'' and the THGEM holes is proportional to detector's gain. With an induction field of \SI{0.5}{\kilo\volt\per\centi\meter}, the rate of charging-up of the bottom rim is lower compared to the charge accumulation rate on the electrode's walls, therefore the sum of both components result in a decrease of a gain on the short timescale. While charging-up of the inner surface of the holes, and that of the bottom rim is responsible for the short-term stabilization time, the ``\textit{top rim}'' charging up is associated with the long-term stabilization time. The difference between long-term and short-term stabilization time is found to be two-fold in an electrode with \SI{0.1}{\milli\metre} hole-rim, and is consistent with the difference between long and the short time-scales obtained in the simulations. Further investigation on the ratio between the short and the long timescales will be discussed in the next section.

\subsubsection{Charging up of the ``\textit{top rim}''}
\begin{figure}[htbp]
\centering
\includegraphics[width=0.8\textwidth,origin=c,angle=0]{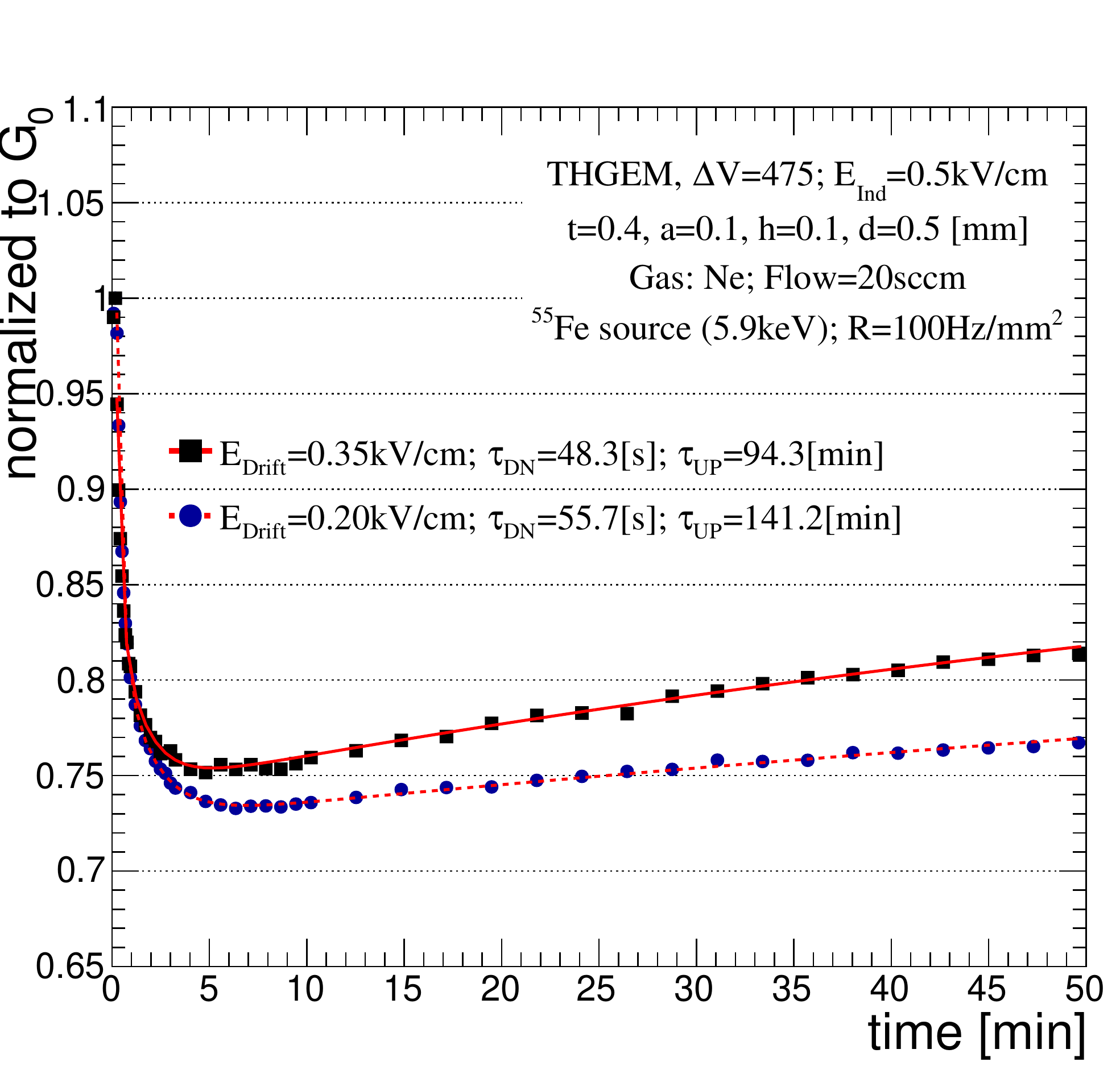}
\caption{\label{fig:toprim} Measured gain stabilization over time in Ne, at an X-ray rate of \SI{100}{\hertz\per\milli\metre\squared} for $E_\text{Drift}$=\SI{0.35}{\kilo\volt\per\centi\meter} (blue) and $E_\text{Drift}$=\SI{0.2}{\kilo\volt\per\centi\meter} (black). The red lines are fit results of the measurement to a triple-exponential function~\eqref{eq:exp3fit}.}
\end{figure}

The drift field permits drifting the primary electron cloud into the THGEM holes. With an increased drift field at fixed electrode voltage, the electron transfer efficiency (ETE) of the primary electrons decreases. It is a known fact that at higher fields a growing fraction of the electrons are collected on the top THGEM electrode, rather than entering the holes~\cite{SHALEM2006475}. The maximum gain in Ne was achieved at drift fields of $\sim$ \SI{0.2}{\kilo\volt\per\centi\meter}, under a drift field of  \SI{0.35}{\kilo\volt\per\centi\meter}, the pulse-height is affected by the strongly decreasing ETE, and a fraction of the primary charge that is not focused into THGEM holes being collected on the top electrode or accumulated on the ``\textit{top-rim}''. Figure~\ref{fig:toprim} depicts the long-term gain stabilization at drift fields of \SI{0.2}{\kilo\volt\per\centi\meter} and \SI{0.35}{\kilo\volt\per\centi\meter}.

While the effective gain of the multiplier is a product of ETE and the detector's gain ($G_\text{eff}=G\times\left(\text{ETE}\right)$), the ratio of the long- to short-term stabilization time is proportional to:
\begin{equation}
\label{eq:ratioLongShort}
\frac{\tau_\text{UP}}{\tau_\text{DN}} \propto \frac{G\times\text{ETE}}{\left(1-\text{ETE}\right)}
\end{equation}
For higher drift-field values, two parameters are changing the bare gain of the detector increases, and the ETE decreases. The measured effective gain decreased by an amount of 25\% with the increased drift-field values from 0.2 to \SI{0.35}{\kilo\volt\per\centi\meter}. This will result in higher increase of the ratio between long-term and short-term characteristic time scales as it shown in figure~\ref{fig:toprim}. However, the loss of primary electrons due to reduced ETE will affect the energy resolution. According to the simulation results (figure~\ref{fig:slices}) a few \SI{}{\pico\coulomb} of charge are needed to charge up completely the ``\textit{top rim}'', then the long-term stabilization time will be given by:
\begin{equation}
\label{eq:probTop}
Q_\text{top-rim} = f_\text{top}\times n_p\times q_e\times R \times\tau_\text{UP} 
\end{equation}
Where $f_\text{top}$ is the fraction of the primary electron to end up on the ``\textit{top rim}'', and the $Q_\text{top-rim}$ is the total charge accumulated on the top-rim (usually of few \SI{}{\pico\coulomb}).

\subsubsection{Charging up of the ``\textit{bottom rim}''}
The charging-up rate of the ``\textit{bottom rim}'' is similar to that of the THGEM holes; therefore, the gain increase induced by the charged ``\textit{bottom rim}'' is suppressed by its decrease due to charging up of the THGEM-hole walls. The typical value of the induction field, \SI{1}{\kilo\volt\per\centi\meter}, assures large fraction of the induced signal to be recorded by the detector's anode (up to 100\%). A reverse or a small induction field draws avalanche electrons to the bottom-THGEM face. At the vicinity of the detector's ``\textit{bottom rim}'', the electric field causes additional multiplication of the avalanche electrons; therefore, a fraction of the generated avalanche ions is usually attached to the ``\textit{bottom rim}''. 
  
\begin{figure}[htbp]
\centering
\includegraphics[width=0.8\textwidth,origin=c,angle=0]{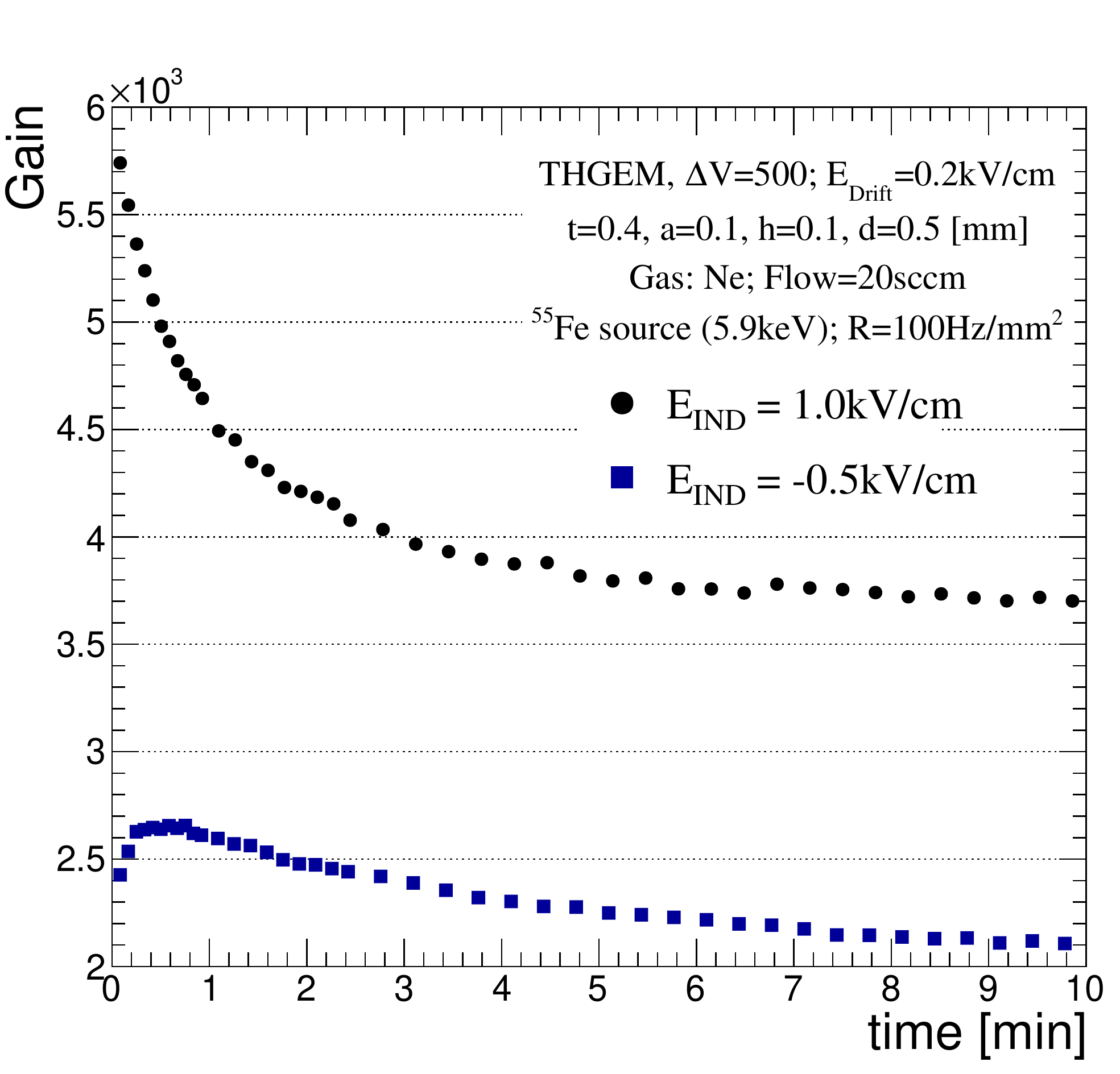}
\caption{\label{fig:bottomrim} Measured THGEM gain stabilization over time in Ne, at an X-ray rate of \SI{100}{\hertz\per\milli\metre\squared} for $E_\text{IND}$=\SI{1}{\kilo\volt\per\centi\meter} (black) and reverse induction field $E_\text{IND}$=\SI{-0.5}{\kilo\volt\per\centi\meter} (blue).}
\end{figure}

Gain stabilization was measured for a reverse induction field, where all the avalanche electrons are collected at the bottom-THGEM electrode, compared to one with $E_\text{IND}$=\SI{1}{\kilo\volt\per\centi\meter}, where most of the avalanche electrons are drawn towards the anode, as shown in figure~\ref{fig:bottomrim}.
With reversed induction field, avalanche electrons that undergo additional multiplication at the vicinity of the ``\textit{bottom rim}'' increase the rate of the bottom-rim charging up (compared to that of THGEM holes). In this case, the timescale of charging up of the ``\textit{bottom rim}'' is higher, which result in an initial increase of the gain. The ``\textit{bottom rim}'' is fully charged, while the detector's walls continue to charge up, and after initial increase of the gain, the gain drop in the timescale of several minutes. As one can note the stable-gain value is higher for the induction field of \SI{1}{\kilo\volt\per\centi\meter} At high induction (or transfer) fields, the strength of the electric field just outside the hole is higher; this results in an increase of the detector gain, due to additional small charge multiplication in the induction gap~\cite{1748-0221-8-06-P06004}.

\section{Conclusions}\label{sec:conclusions}
Gain stabilization processes of THGEM-based detectors where measured following the methodology developed in~\cite{1748-0221-12-09-P09036}. The measurement results were systematically compared to the simulated ones, based on the model suggested in~\cite{Correia:2017qdx}. This model attributes the gain variations to the charging up of the detector's insulator substrate at the initial operation period. These charging-up transients should not affect the detector's long-term operation.

While modeling the charging-up process using a single-exponential function could fit the simulation results, the experimental data was better fitted to a double-exponential model (figure~\ref{fig:simcompare}). The difference in the shape of the respective graphs in figure~\ref{fig:simcompare} could have originated from experimental conditions, electrode's inhomogeneity or small amounts of gas impurities -- all capable of affecting the gain stabilization~\cite{1748-0221-10-09-P09020,1748-0221-12-09-P09036}. It was found that charging-up processes can be quantitatively modeled through basic parameters such as ``characteristic time'' -- the time it takes the gain to reach half of its total variation, or ``characteristic charge'' -- the total amount of charge passing through the THGEM holes during the relaxation period~\eqref{eq:qtot}. The later could also indicate the rate of stabilization. We found a linear dependence between gain-stabilization time and irradiation rate, supporting the argument that the main mechanism affecting THGEM-based detector's gain evolution is the charging-up of the insulator surfaces. We also show that detector stabilization time depends linearly on the inverse of the detector's gain.

The effect of the hole-rim on the detector's gain stabilization was systematically investigated;  it was found that the ``\textit{top rim}'' introduces a long-term gain-stabilization time, typically two-fold different compared to the short one. Both simulation and experimental data indicate that the long-term stabilization time can be order of magnitude larger than the short-term one. For detector's electrodes having hole-rims, a short-term stabilization time appears as an initial gain decrease, while the long-term one is reflected by a slow increase in the detector's gain. It has also been shown that the rate of charging-up of the ``\textit{top rim}'' depends on the focusing of the primary-electrons cloud into the THGEM's holes, and it can vary by an order of magnitude in time at a measured rate of \SI{100}{\hertz\per\milli\metre\squared}. The charging-up of the ``\textit{bottom rim}'' reduces the total gain drop due to an accumulation of avalanche-ions on the ``\textit{bottom rim}'' surface. In figure~\ref{fig:rimapart}, simulated results show an exponential decrease in both cases: when the ``\textit{bottom rim}'' is charged-up and when the charges are not allowed to accumulate on the ``\textit{bottom rim}'' surface. The simulations suggest that the charging-up of the ``\textit{bottom rim}'' increases the gain, as confirmed by the measurements with reversed induction field. Once reversed, the charging-up rate of the ``\textit{bottom rim}'' became higher than that of the hole's walls, resulting in an initial gain increase (figure~\ref{fig:bottomrim}). The measurements show that the charging-up transients relate only to initial detector's operation period, and were not observed during long-term operation.

\acknowledgments
This research was supported in part by the I-CORE Program of the Israel Planning and Budgeting Committee, the Nella and Leon Benoziyo Center for High Energy Physics at the Weizmann Institute of Science, Grant \textnumero 712482 from the Israeli Science Foundation (ISF) and by CERN/FIS-INS/0025/2017 project through COMPETE, FEDER and FCT (Lisbon) programs.

\bibliographystyle{JHEP}
\bibliography{charging}

\end{document}